\documentclass{kluwer}    

\usepackage{psfig}

\newdisplay{guess}{Conjecture}

\begin{document}                                                                                   
\begin{article}
\begin{opening}         

\title{
Hierarchical learning in polynomial Support Vector \\
Machines 
}

\author{Sebastian \surname{Risau-Gusman} and Mirta B. \surname{Gordon}}
\runningauthor{Sebastian Risau-Gusman and Mirta B. Gordon}   
\runningtitle{Generalization properties of finite size polynomial Support 
Vector Machines}
\institute{DRFMC/SPSMS CEA Grenoble, 17 av. des Martyrs\\  
38054 Grenoble Cedex 09, France  
}  

\date{March 14, 2000}

\begin{abstract}
We study the typical properties of polynomial Support 
Vector Machines within a Statistical Mechanics approach 
that allows us to analyze the effect of different 
normalizations of the features. If the 
normalization is adecuately chosen, there is a hierarchical
learning of features of increasing order as a function 
of the training set size.
\end{abstract}

\end{opening}       
  
\section{Introduction}

The theoretical basis of learning from examples relies on 
the possibility of bounding the generalization error, which 
is the probability that the trained machine makes an error 
on a new pattern. However, the rigorous bounds deduced with methods 
of statistical theory \cite{Vapnik}, which hold for the 
worst case, turn out to be too pessimistic in most real 
world applications. Bounds to the {\it estimator} of the 
generalization error, obtained by the leave-one-out technique, 
are closer to experimental results \cite{VCh}. This estimator 
is obtained through averaging the classification error on one 
pattern, when learning was achieved by removing that pattern 
from the training set. Another  approach to learning theory, 
heralded more than a decade ago by the pioneering work of 
E. Gardner on perceptrons \cite{GD}, strives to determine 
analytically the {\it typical} properties of the learning machine 
under somewhat restrictive hypothesis. This is done using 
methods from Statistical Physics, developed to study the 
properties of large, disordered physical systems. The basic 
assumption of this approach is that average quantities are 
representative of the machine's properties, with a probability 
that tends to the unity as the system's size goes to infinity, 
a limit usually called the {\it thermodynamic} (TD) limit. Thus, 
in this limit, the leave-one-out estimators and the Statistical 
Mechanics results are expected to converge to the same values. The 
input patterns distribution plays the role of the disorder over 
which averages are taken, the size of the system being the input 
space dimension. In the TD limit, some control parameters, 
like the training set size relative to the input space 
dimension, are kept fixed. This enables to deduce typical 
properties for small relative training set sizes, in contrast 
to the bounds provided by the statistical theories, which are 
generally valid for sufficiently large training sets. 
In this paper we are able to get deeper insight on the typical 
properties of Support Vector Machines (SVMs) by also keeping 
fixed other characteristic quantities.

The typical properties of polynomial SVMs have been studied in 
two recent papers \cite{DOS,BG}. Both consider SVMs in which the 
input vectors ${\bf x} \in \Re^n$ are mapped onto quadratic features 
$\bf \Phi$ using the {\it normalized} mapping \cite{DOS}

\begin{equation}
{\bf \Phi}_N({\bf x})=({\bf x}, x_1{\bf x}/\sqrt{n}, 
x_2{\bf x}/\sqrt{n}, \cdots, x_n{\bf x}/\sqrt{n}),
\end{equation}
 
\noindent and the {\it non-normalized} mapping \cite{BG} 

\begin{equation}
{\bf \Phi}_{NN}({\bf x})=({\bf x}, x_1{\bf x}, x_2{\bf x}, \cdots, x_{k}{\bf x}).
\end{equation}
 
\noindent respectively. The latter was studied as a function 
of $k$, the number of quadratic features. For $k=n$ the dimension 
of both feature spaces is the same. They correspond to the quadratic 
kernels 

\begin{equation}
\label{eq.kernel}
K({\bf x}, {\bf y})={\bf x} \cdot {\bf y} \, \, (1+a \, {\bf x} \cdot {\bf y}),
\end{equation}

\noindent with $a=1/\sqrt{n}$ for ${\bf \Phi}_N$ and $a=1$ for 
${\bf \Phi}_{NN}$. In spite of the seemingly innocuous differences 
between the models, their properties are very different. 

If the rule to be learned is a linear separation in input space, 
the generalization error $\epsilon_g$ corresponding to the non-normalized 
mapping ${\bf \Phi}_{NN}$ is much larger than the 
Maximal Margin Hyperplane (MMH) solution of a simple perceptron 
in input space (i.e. with no added 
features). The latter corresponds to a linear SVM, which is the 
smallest SVM able to generalize this rule. This difference between 
the quadratic and the linear SVMs increases dramatically with $k$, 
the number of quadratic features included. On the other hand, the 
generalization error corresponding to the normalized mapping 
${\bf \Phi}_N$ is only slightly larger than that of the simple 
MMH perceptron. In the case of learning quadratic separations, 
the generalization error $\epsilon_g$ of the normalized mapping 
exhibits a very interesting behaviour \cite{DOS,YO}: if the 
number of training patterns scales with $n$, the 
dimension of the linear subspace, $\epsilon_g$ decreases up 
to a finite asymptotic lower bound, and it only vanishes 
asymptotically if the number of patterns scales proportionally 
to $n^2$. The generalization error of higher order polynomial SVMs 
using the normalized mapping also presents different scaling 
regimes \cite{DOS}. 

In order to understand these results, it is useful to consider 
the feature space of the polynomial SVMs as the direct product 
of several subspaces, each one spanned by all the monomials (of the 
same degree) that can be formed with the input components. 
The number of subspaces is given by the degree of the polynomial. 
The dimension of each subspace, equal to the number of different 
monomials, grows polynomially with $n$, the input space 
dimension. For example, in quadratic SVM's there are two 
subspaces, one corresponding to the $n$ linear features and the 
other to the $n^2$ quadratic features. In the case of the 
normalized mapping, if the training set size 
scales with the dimension of one particular subspace, only the 
features belonging to this and to the lower order subspaces 
contribute effectively to learning \cite{DOS}. As a result, 
the generalization error decreases asymptotically 
to a lower bound. The latter is smaller the higher the particular 
subspace dimension, and goes to zero when the number of training 
patterns is proportional to the dimension of the highest 
dimensional subspace. We call this behaviour {\it hierarchical 
learning}.

The only difference between the mappings ${\bf \Phi}_N$ and 
${\bf \Phi}_{NN}$ is that the quadratic features in ${\bf \Phi}_N$ 
are squeezed by a factor $a=1/\sqrt{n}$ with respect to those of 
${\bf \Phi}_{NN}$. This normalization is very sensitive to the 
TD limit. In particular, at finite $n$, the hierarchical learning 
behaviour of the generalization error, sharply defined in the TD 
limit, is expected to give raise to crossovers between successive 
regimes as a function of the number of training patterns, in which 
features belonging to increasingly higher order subspaces are learned. 

The aim of the present paper is to clarify to what extent the 
qualitative differences between the normalized and the non-normalized 
mappings are still present in polynomial SVMs working in {\it finite} 
dimension, and to characterize the signature of the hierarchical 
learning. Within our approach, given the input space dimension $n$ 
and the mapping, the polynomial SVM is characterized by two kinds 
of quantities, the {\it inflation factor} and the {\it feature's 
variance} in each subspace. The former is given by the corresponding 
subspace dimension relative to the input space dimension. The latter 
is proportional to the normalizing factor $a$; depending on its 
value, the features distribution may be highly {\it anisotropic} 
in the sense that in different subspaces the corresponding features 
have different variances. 

Using the tools of Statistical Mechanics, we determine the 
typical properties of SVMs of finite inflation factors and 
features variances, as a function of $\alpha \equiv \ell/n$, 
where $\ell$ is the the number of training patterns. This is 
different from the approaches of  \cite{DOS,YO} who considered 
separately two different scalings for $\ell$, namely 
$\ell \propto n$ and $\ell \propto n^2$. Our results give some 
insight on the inner workings of finite SVMs, given their 
inflation factors and variances. The different scaling regimes 
of the generalization error as a function of $\alpha$ become 
a crossover, as expected, but more interestingly, its steepness 
depends not only on the particular SVM considered, but also on 
the complexity of the rule to be learnt. The previous results 
\cite{DOS,BG} are recovered by performing the suitable 
limits. Our analytical results are supported by the excellent 
agreement with the numerical simulations. 

The paper is organized as follows: in the next section we 
present our model. A short introduction to the Statistical 
Mechanics approach, with its application to our analysis of 
the quadratic SVMs is presented in section \ref{sec.SM}, the 
details being left to the Appendix. The analytic predictions 
are described and compared with numerical simulations in 
\ref{sec.results}. The results are discussed and generalized 
in section \ref{sec.discussion}. The conclusion is left to 
section \ref{sec.conclusion}.

\section{The Model}
\label{sec.model}

We consider the problem of learning a binary classification task from examples with a SVM in polynomial feature spaces. 
The learning set contains $\ell$ patterns $({\bf x}_i, y_i)$ ($i=1,\cdots,\ell$) where ${\bf x}_i$ is an input 
vector in the $n$-dimensional input space, and 
$y_i \in \{-1,1\}$ is its class. We assume that the 
components $x_{i \nu}$ ($\nu=1, \cdots, n$) are independent 
identically distributed gaussian random variables conveniently standardized, so that they have zero-mean and unit variance:

\begin{equation}
\label{eq.pdex}  
P({\bf x}) = \prod_{\nu=1}^{n} \frac{1}{\sqrt{2 \pi}} \exp \left( - \frac{ x_{\nu}^2} 
{2} \right).  
\end{equation} 

\noindent These input vectors are mapped to a higher 
dimensional space (the feature space) wherein the machine 
looks for the MMH. In the following we concentrate on 
quadratic feature spaces, although our conclusions are 
more general, and may be applied to higher order 
polynomial SVMs, as discussed in section \ref{sec.discussion}. 
The mappings ${\bf \Phi}_{NN}({\bf x})=({\bf x},x_1{\bf x}, 
x_2{\bf x}, \cdots, x_n{\bf x})$ and  ${\bf \Phi}_N
({\bf x})=({\bf x}, x_1{\bf x}/\sqrt{n},  x_2{\bf x}/\sqrt{n}, 
\cdots, x_n{\bf x}/\sqrt{n})$ are particular instances of 
mappings of the form ${\bf \Phi} ({\bf x})= (\phi_1, \cdots, 
\phi_n, \phi_{11},  \phi_{12},\cdots, \phi_{nn})$ with 
$\phi_\nu=x_\nu$, and $\phi_{\nu \mu}=a \, x_\nu x_\mu$. 
$a$ is the {\it normalizing factor} of the quadratic 
components: $a=1$ for mapping ${\bf \Phi}_{NN}$ and 
$a=1/\sqrt{n}$ for ${\bf \Phi}_{N}$. The patterns probability 
distribution in feature-space is:

\begin{equation}
\label{eq.pdephi}
P\left({\mbox{$\bf \Phi$}}\right) = \int \, \prod_{\nu=1}^{n} 
\frac{dx_\nu}{\sqrt{2 \pi}} \exp \left( - \frac{ x_\nu^2} {2} \right) 
\delta(\phi_\nu - x_\nu)
\prod_{\mu=1}^{n} \delta \left(\phi_{\nu \mu}-a \, x_\nu x_\mu \right). 
\end{equation}
 
\noindent Clearly, the components of ${\bf \Phi}$ are not 
independent random variables. For example, a number $O(n^3)$ 
of triplets of the form $\phi_{\nu \mu} 
\phi_{\mu \tau} \phi_{\tau \nu}$ have positive correlations. These 
contribute to the third order moments, which should 
vanish if the features were gaussian. Moreover, the 
fourth order connected correlations \cite{Monasson} 
do not vanish in the thermodynamic limit. Nevertheless, 
in the following we will neglect these and 
higher order connected moments. This approximation, 
used in \cite{BG} and implicit in \cite{DOS}, is 
equivalent to assuming that all the components in 
feature space are independent gaussian variables. 
Then, the only difference between the mappings 
${\bf\Phi}_{N}$ and ${\bf\Phi}_{NN}$ lies in the variance of 
the quadratic components distribution. The results 
obtained using this simplification are in excellent 
agreement with the numerical tests described in 
the next section.

Since, due to the symmetry of the transformation, only 
$n(n+1)/2$ among the $n^2$ quadratic features are different, 
hereafter we restrict the feature space 
and only consider the non redundant components, that 
we denote ${\bf z}=({\bf z}^u,
{\bf z}^\sigma)$. 
The first $n$ components ${\bf z}^u = 
(z_1, \cdots, z_n)$ hereafter called {\it u-components}, 
represent the original input pattern of unit variance, 
lying in the linear subspace of the feature
space. The remaining components 
${\bf z}^\sigma= (z_{n+1}, \cdots, z_{\tilde n})$ 
stand for the {\it non redundant} quadratic features, of 
variance $\sigma$, hereafter called $\sigma$-{\it components}. 
${\tilde n} = n(1+\Delta)$ is the dimension  of the restricted 
feature space, where the {\it inflation ratio} 
$\Delta$ is the relative number of non-redundant quadratic 
features per input space dimension. The quadratic mapping 
has $\Delta=(n+1)/2$. 

According to the preceding discussion, we assume that 
learning $n$-dimensional patterns selected with the 
isotropic distribution (\ref{eq.pdex}) with a 
quadratic SVM is equivalent to learning the MMH with 
a simple perceptron in an ${\tilde n}$-dimensional space 
where the patterns are drawn using the following 
anisotropic distribution, 

\begin{equation}
\label{eq.pdexi}
P\left(\bf z\right) = 
\frac{1}{(2 \pi)^{n/2}} \exp \left[-\frac{({\bf z}^u)^2}{2} \right] \; 
\frac{1}{(2 \pi \sigma^2)^{n \Delta/2}} \exp \left[-\frac{
({\bf z}^\sigma)^2}{2 \sigma^2} \right],
\end{equation} 

\noindent where

\begin{eqnarray}
\label{eq.delta}
\Delta &=& \frac{n + 1}{2}, \\
\label{eq.sigma2}
\sigma^2 &=& \frac{n a^2}{\Delta}
\end{eqnarray}

\noindent with $a$ the normalization factor 
in (\ref{eq.kernel}). The second moment of the {\it u}-features 
is $\langle ({{\bf z}}^u)^2 
\rangle = n$ and that of the $\sigma$-features 
is $\langle ({{\bf z}}^\sigma)^2 
\rangle = n \Delta \sigma^2$. If $\sigma^2 \Delta=1$, 
we get $\langle ({{\bf z}}^\sigma)^2 \rangle 
= \langle ({{\bf z}}^u)^2 \rangle$, which 
is the relation satisfied by the normalized mapping 
considered in \cite{DOS}. The non-normalized mapping 
satisfies $\sigma^2 \Delta=n$. 

\section{Statistical Mechanics}
\label{sec.SM}

Statistical Mechanics is the branch of Physics that 
deals with the properties of systems with many degrees of 
freedom, characterized by their energy. In our case, this 
is the cost function $E({\bf w}; {\mathcal L}_\ell)$. It 
depends on the weights ${\bf w} \in \Re^{\tilde n}$, whose 
$\tilde n = n(1+\Delta)$ components are the system's degrees 
of freedom, and on the set of training patterns 
$\mathcal{L}_\ell$, which are random variables selected 
with distribution (\ref{eq.pdexi}). The equilibrium properties 
at temperature $T \equiv \beta^{-1}$ are deduced through 
averages over the Gibbs distribution:

\begin{equation} 
P_\beta({\bf w};{\mathcal L}_\ell) = \frac{1}{Z_\beta({\mathcal L}_\ell)} \exp(-\beta E(\bf w;{\mathcal L}_\ell)),  
\label{eq.Gibbs}
\end{equation} 

\noindent $Z_\beta({\mathcal L}_\ell)$ is a normalization 
constant called partition function, defined through

\begin{equation}
\label{eq.partition}
Z_\beta({\mathcal L}_\ell)=\int \exp [-\beta E({\bf w}; {\mathcal L}_\ell) ] \; p({\bf w}) \, d{\bf w},
\end{equation}

\noindent where $p({\bf w}) \, d{\bf w}$ is a measure in 
the weights' space. When $\beta$ is very large, only weights 
$\bf w$ with energies close to the minimum have significant 
probability, and contribute to the the integral in 
(\ref{eq.partition}). In the limit $\beta \rightarrow \infty$, 
only those that have minimal energy have non-vanishing probability. 

The fraction of training errors corresponding to the minimal cost, 
or any other intensive property of the system, may be deduced from 
the so called free energy $f_\beta({\mathcal L}_\ell) \equiv -\ln 
Z_\beta({\mathcal L}_\ell)/(\beta n)$ (or through its derivatives 
with respect to $\beta$), in the limit $\beta \rightarrow \infty$. 
These quantities are random variables, as they depend on the 
realization of ${\mathcal L}_\ell$. In the TD limit 
$n \rightarrow \infty$, their variance is expected to vanish 
like $1/\sqrt{n}$, which means that all the training sets of the 
same size are expected to endow the system with the same intensive 
{\it typical} properties with probability one. This hypothesis, 
supported in particular by the good agreement between the 
predictions and the simulations on perceptrons, has been 
recently shown to hold on theoretical grounds, at least 
for some quantities \cite{Tal}. As $f_\beta({\mathcal L}_\ell)$ 
is equal with probability $1$ for all the training sets, we can 
get rid of the dependance on $\mathcal{L}_\ell$ by taking mean 
values over all the possible training sets. This non-trivial 
average is done using a sophisticated technique, known as the 
replica trick, developed for the study of disordered magnetic systems \cite{MPV}, which uses the identity:

\begin{equation}
\label{eq.lnZ}
\overline{\ln Z}=\lim_{m \rightarrow 0} \frac{\overline{Z^m}-1}{m}
\end{equation}

\noindent where the overline stands for the average over 
the training sets. The average of $\ln Z$ has been transformed 
into that of averaging the partition functions of $m$ replicated 
systems. In order to obtain non-trivial results, one takes also $\ell \rightarrow \infty$ with the number of patterns per input 
space dimension $\alpha \equiv \ell/n$ constant. 
If we want to compare the results given by this approach 
with the corresponding quantities of finite machines, $\alpha$
is a relevant parameter. 

This formalism has been succesfully applied to 
the analysis of neural networks \cite{GD,SST,WR}. 
In particular, an interesting property is the 
generalization error of the trained machine, which 
in principle depends on the training set. The above 
assumptions amount to saying that in the TD limit, 
the probability that the generalization error is $\epsilon_g$
is a delta peak centered at the typical value, 
of zero variance. At finite input 
dimension $n$, this probability distribution widens, 
and its mean value is shifted. Both these corrections 
are of order $O(1/\sqrt{n})$. This behaviour, 
numerically verified within several learning scenarios 
\cite{BG-bayesian,NF,SU}, shows that the predictions of 
the statistical mechanics approach are better for 
larger $n$. 

In our case, ${\mathcal L}_\ell
=\{({\bf z}_i, y_i)\}_
{i=1,\cdots,\ell}$ is the set of training patterns 
in feature space, with ${\bf z}_i = ({\bf z}^u_i,{\bf z}^
\sigma_i)$ drawn with probability (\ref{eq.pdexi}), the 
$y_i \in \{-1,+1\}$ being the corresponding classes. The 
natural cost function $E$ is:

\begin{equation} 
\label{eq.cost} 
E({\bf w}, \kappa; {\mathcal L}_\ell)=\sum_{i=1}^\ell \Theta(\kappa-\gamma_i) \;
\end{equation}  

\noindent where $\Theta$ is the Heaviside function, 
$\gamma_i = y_i {\bf z}_i 
\cdot \bf w / \sqrt{{\bf w} \cdot {\bf w}}$ is the 
{\it margin} of pattern ${\bf z}_i$, 
and $\kappa$ is the minimal margin, the smallest 
allowed distance between the hyperplane and the 
training patterns. Energy (\ref{eq.cost}) is thus the number 
of patterns with margin smaller than $\kappa$. The perceptron 
with weights corresponding to a vanishing cost that maximizes 
$\kappa$ implements the Maximal Margin Hyperplane (MMH). 

The properties of the MMH for the most ``natural'' 
pattern distribution, an isotropic gaussian, have 
been thoroughly studied \cite{OKKN,GG}. 
The case of a single anisotropy axis has also been 
investigated \cite{MBS}. In our model, the properties 
of a polynomial SVM are those of a MMH perceptron with 
its inputs lying in the feature space, drawn from 
the anisotropic gaussian (\ref{eq.pdexi}), with macroscopic 
subsets of components having different variances. Since 
the rules to be inferred are assumed to be linear 
separations in feature space, we represent them by 
the weights ${\bf w}_*=({\bf w}^u_*,{\bf w}^{\sigma}_*)$ 
of a {\it teacher perceptron}, so that the class of 
the patterns is $y ={\rm sign}({\bf z} 
\cdot {\bf w}_*)$. Without any loss of generality we 
consider normalized teachers: ${\bf w}_* \cdot {\bf w}_* 
= \tilde n$. For the students weights 
${\bf w}=({\bf w}^u,{\bf w}^{\sigma})$ we take the same 
normalization: ${\bf w} \cdot {\bf w} \equiv 
{\bf w}^u \cdot {\bf w}^u + 
{\bf w}^{\sigma} \cdot {\bf w}^{\sigma}=\tilde n$ .

Within our model, the values of $\Delta$ and $\sigma$ 
depend on the particular (finite size) SVM we want to analyze, 
through equations (\ref{eq.delta}) and (\ref{eq.sigma2}). 
The main difference with respect to previous work 
\cite{DOS,BG} is that we can make predictions 
that take into account the mapping normalization 
and some finite-size characteristics of the SVM.

We calculate the typical properties of a particular SVM 
as follows: we fix $\Delta$ and $\sigma$ as explained above, 
and calculate the averages using distributions (\ref{eq.Gibbs}) 
and (\ref{eq.pdexi}), in the limits $\beta \rightarrow \infty$, 
$n \rightarrow \infty$, $\ell \rightarrow \infty$ with 
$\ell/n \equiv \alpha$ constant. A similar procedure has 
been used in a toy model \cite{R-GG} in the context of 
Gibbs learning. To obtain the properties of the MMH, we 
look for the maximal value of $\kappa$ with vanishing 
cost (\ref{eq.cost}). The details of the calculations 
are left to the Appendix. In the following we describe 
the main results.

It turns out that the typical properties of the SVM 
can be expressed as a function of the (normalized) squared norm 
of the teacher weights in $\sigma$-subspace,

\begin{equation}
\label{eq.Q*}
Q_* = \frac{{\bf w}_*^{\sigma} \cdot {\bf w}_*^{\sigma}}
{\tilde n}, 
\end{equation}

\noindent and the following averages (also called order 
parameters, in physics):

\begin{eqnarray} 
\label{eq.Q}
Q &=& \frac{\overline{\langle {\bf w}^u \cdot {\bf w}^u \rangle}}{\tilde n},
\\
R^u &=& \frac{\overline{\langle {\bf w}^u \cdot {\bf w}_*^u \rangle}}{\tilde n \sqrt{(1-Q)(1-Q_*)} } , 
\\
\label{eq.Rsigma}
R^\sigma &=& \frac{\overline{\langle {\bf w}^{\sigma} \cdot {\bf w}_*^{\sigma}} \rangle}{\tilde n  \sqrt{Q \, Q_*}},
\end{eqnarray}

\noindent where $\langle \cdots \rangle$ represents the 
averaging over distribution (\ref{eq.Gibbs}). In the 
limit $\beta \rightarrow \infty$, this last averaging 
is trivial because the MMH is unique. $Q$ is 
the squared norm of the student weights in 
$\sigma$-subspace. $R^u$ is the 
average of the overlap between the weights of the teacher 
and the student in the $u$-subspace. $R^{\sigma}$ is the 
corresponding overlap in the $\sigma$-subspace. Notice that 
the denominators ensure that both R's are conveniently 
normalized, that is, that their values lie between 0 and 1. 

The generalization error, defined as the probability that 
the machine with weights $\bf w$ makes an error on an unknown
pattern, 

\begin{equation}
E_g = \int \Theta \left[- ({\bf z} \cdot {\bf w_*}) ({\bf z} \cdot {\bf w})\right] 
\, P ({\bf z}) 
\, d {\bf z},
\end{equation}

\noindent can be expressed in terms of the order parameters. After 
a straightforward but cumbersome calculation, we find that its typical 
value can be written as:

\begin{eqnarray}
\label{eq.eg}
\epsilon_g \equiv \overline{\langle E_g \rangle}= \frac{1}{\pi} \arccos(R) \;
\end{eqnarray}

\noindent where

\begin{equation}
\label{eq.R}
R = \frac{R^u + \sqrt{\Delta^\sigma \Delta_*^\sigma} \, R^\sigma} 
{\sqrt{(1+\Delta^\sigma) (1+\Delta_*^\sigma)}} 
\end{equation}

\noindent with $\Delta^\sigma$ and $\Delta_*^\sigma$ given by

\begin{equation}
\label{eq.deltas}
\Delta_{(*)}^\sigma \equiv \frac{\sigma^2 Q_{(*)}}{1-Q_{(*)}}. 
\end{equation}

\noindent The fraction of training patterns that are Support 
Vectors, averaged over all the possible sets of patterns, $\rho_{SV}$, 
is a bound to the leave-one-out estimator of the generalization error 
\cite{Vapnik}. Within our approach it is straightforward 
to calculate it:

\begin{equation}
\label{eq.rho}
\rho_{SV} = 2 \int_{-\infty}^{\frac{\kappa}{\sqrt{(1+\Delta^\sigma) (1-Q)}}} 
H\left(-tR/\sqrt{1-R^2}\right) Dt.
\end{equation}

\noindent where $H(x)=\int_x^\infty Dx$, with 
$Dx=\exp(-x^2/2)/\sqrt{2 \pi}$. 

Thus, the properties 
of the SVM depend on the \emph{teacher} (through 
$Q_*$), on the \emph{inflation factor} of the feature 
space $\Delta$, on the \emph{normalizing factor $a$} 
of the mapping (through $\sigma$) and 
on the \emph{number of patterns per input dimension} of 
the training set ($\alpha$).

\section{Results}
\label{sec.results}

\begin{figure}
\centerline{\psfig{figure=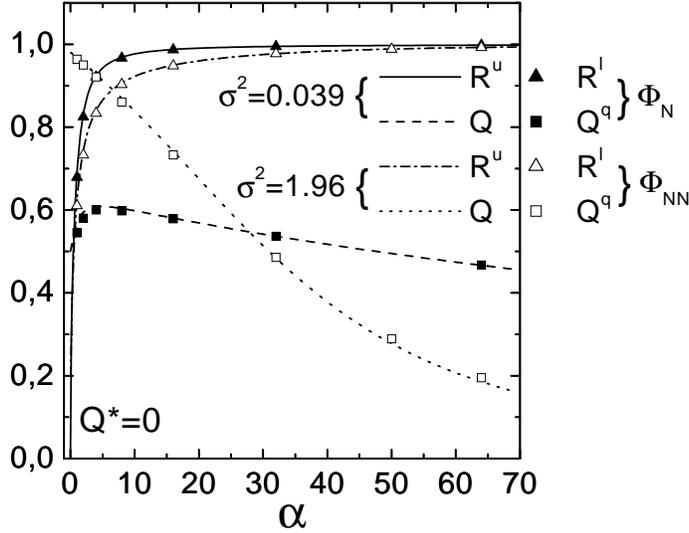,height= 8 cm}}
\caption{Order parameters of SVMs for purely linear 
teacher rules, $Q_*=0$. Symbols are experimental 
results for input space dimension $n=50$, corresponding 
to the two kinds of quadratic mappings, $\Phi_N$ 
with $a=1/\sqrt{n}$ (full symbols) and $\Phi_{NN}$ with 
normalizing factor $a=1$ (open symbols) respectively. 
Error bars are smaller than the symbols. The 
lines are {\it not} fits, but the solutions of the 
Statistical Mechanics equations 
for $\Delta=(n+1)/2$ and $\sigma^2=n a^2/\Delta$ with $n=50$, 
and $a$ corresponding to each mapping.}
\label{fig.Q*0}
\end{figure}

\begin{figure}
\centerline{\psfig{figure=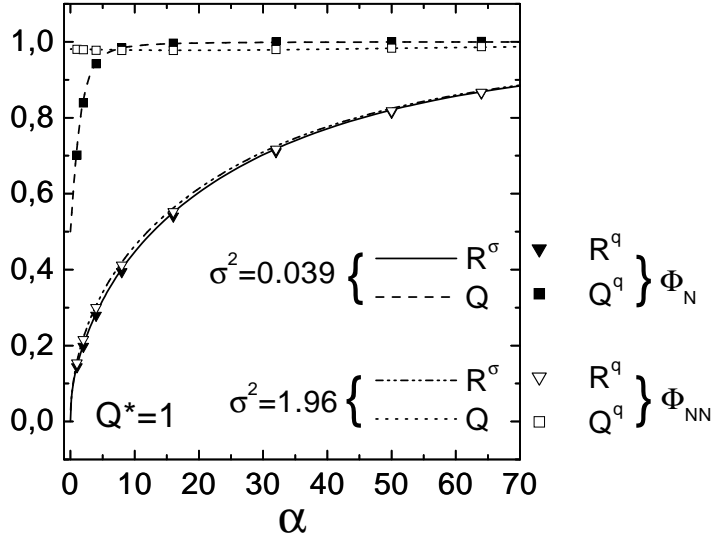,height= 8 cm}}
\caption{Order parameters of SVMs for purely quadratic 
teacher rules, $Q_*=1$. Definitions are the same 
as in figure {\protect {\ref{fig.Q*0}}}.}
\label{fig.Q*1}
\end{figure}

\begin{figure}
\centerline{\psfig{figure=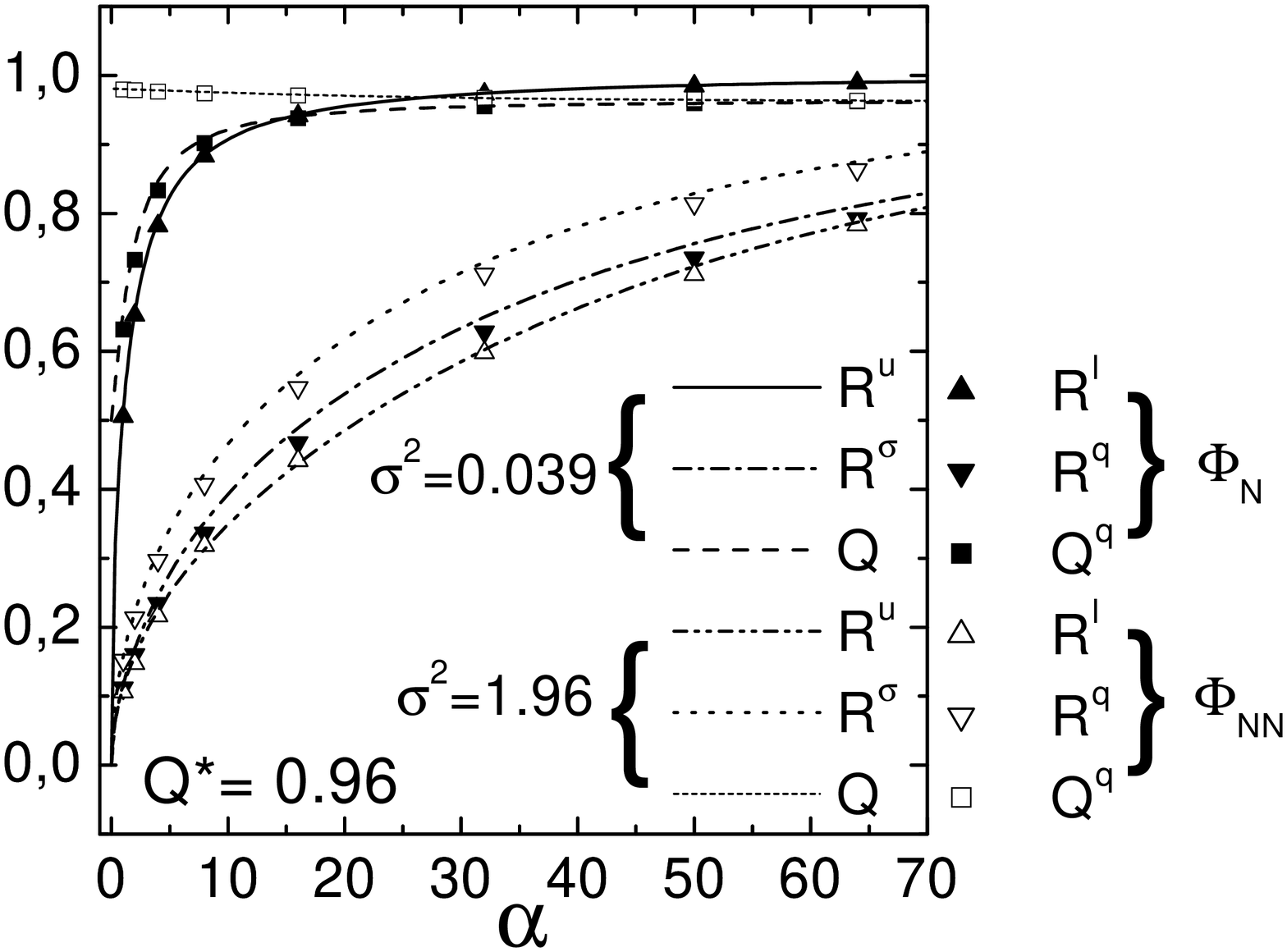,height= 8 cm}}
\caption{Order parameters of SVMs for isotropic 
teacher rules, $Q_{*;\,iso}=\Delta/(1+\Delta)$. Definitions are the same 
as in figure {\protect {\ref{fig.Q*0}}}.}
\label{fig.Q*51/53}
\end{figure}

\begin{figure}
\centerline{\psfig{figure=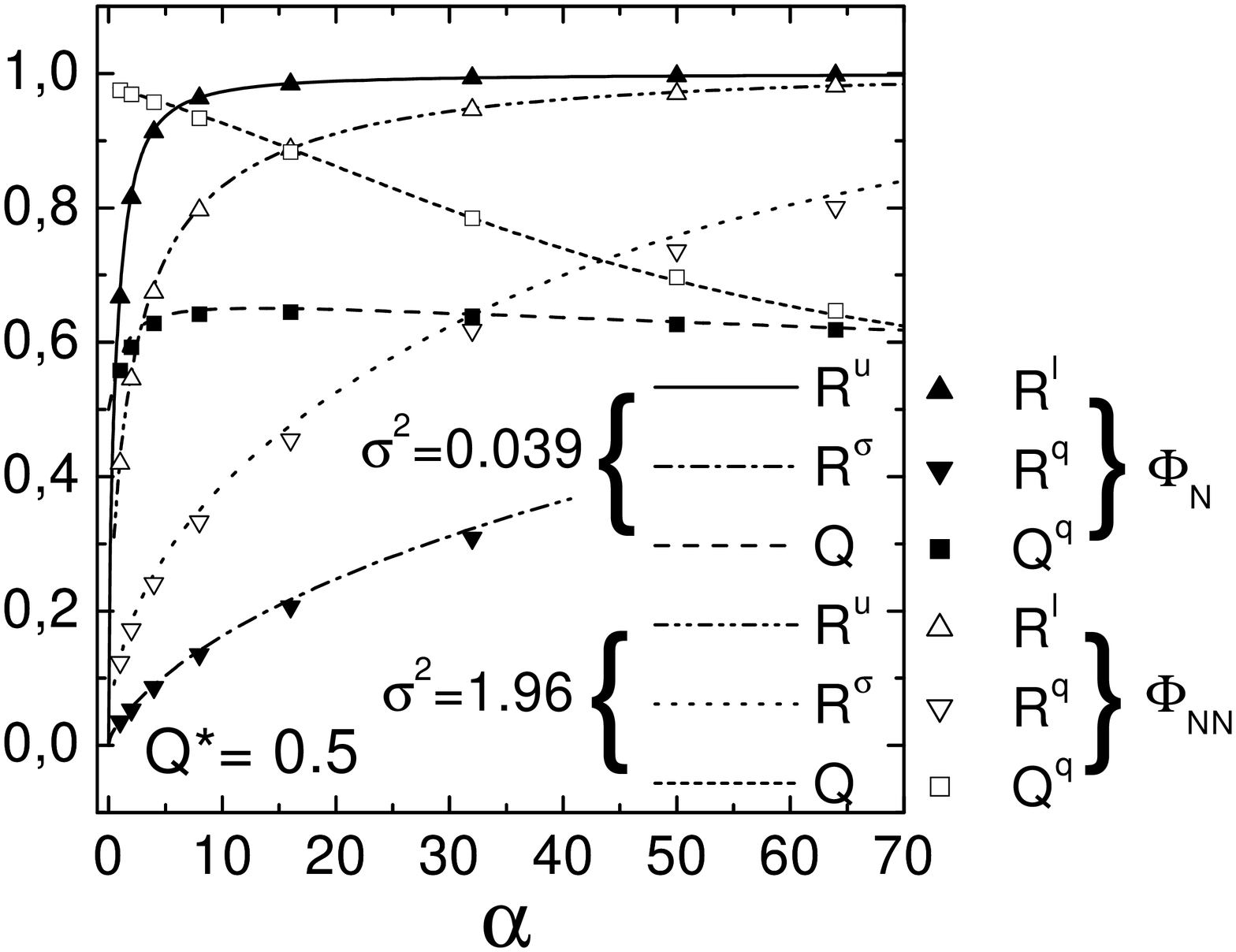,height= 8 cm}}
\caption{Order parameters of SVMs for a general 
teacher rule, $Q_*=0.5$. Definitions are the same 
as in figure {\protect {\ref{fig.Q*0}}}.}
\label{fig.Q*05}
\end{figure}

We describe first the experimental data, obtained 
with quadratic SVMs, using both mappings, ${\bf \Phi}_{NN}$ 
and ${\bf \Phi}_N$, which have normalizing factors $a=1$ 
and $a=1/\sqrt{n}$ respectively, where $n$ is the 
input space dimension. The $\ell= \alpha n$ random input 
examples of each training set were selected with probability 
(\ref{eq.pdex}) and labelled by teachers of normalized weights 
${\bf w}_* \equiv ({\bf w}_*^l,{\bf w}_*^q)$ drawn at random. 
${\bf w}_*^l$ are the $n$ components in the linear 
subspace and ${\bf w}_*^q$ are the $n^2$ components 
in the quadratic subspace.  We do not label the vectors 
like in the previous section to stress the fact 
that the numerical simulations use the 
complete feature space, contrary to the theoretical 
approach where we only consider the non-redundant 
components. Notice that, 
because of the symmetry of the mappings, teachers having 
the same value of the symmetrized weights in the 
quadratic subspace, $(w_{*, \nu \mu}^q+w_{*,\mu\nu}^q)/2$, are 
all equivalent. The teachers are characterized by the proportion of 
(squared) weight components in the quadratic subspace, 
$Q_*={\bf w}_*^q \cdot {\bf w}_*^q/{\bf w}_* \cdot {\bf w}_*$. 
In particular, $Q_*=0$ and $Q_*=1$ correspond to a purely 
linear and a purely quadratic teacher respectively. 
The experimental student weights ${\bf w} \equiv 
({\bf w}^l,{\bf w}^q)$  were obtained by 
solving numerically the dual problem \cite{CV}, using the 
Quadratic Optimizer for Pattern Recognition program \cite{V,AS}, 
that we adapted to the case without threshold treated 
in this paper. We determined $Q^q \equiv {\bf w}^q \cdot {\bf w}^q/n^2$, 
and the normalized overlaps $R^l \equiv {\bf w}^l \cdot {\bf w}_*^l/
\sqrt{ ({\bf w}^l \cdot {\bf w}^l) \, ({\bf w}_*^l \cdot {\bf w}_*^l)}$, 
and $R^q \equiv {\bf w}^q \cdot {\bf w}_*^q/
\sqrt{ ({\bf w}^q \cdot {\bf w}^q) \, ({\bf w}_*^q \cdot {\bf w}_*^q)}$. 
They are represented on figures \ref{fig.Q*0} 
to \ref{fig.Q*05} as a function of $\alpha \equiv \ell/n$, 
using full and open symbols for the mappings ${\bf \Phi}_N$ 
and ${\bf \Phi}_{NN}$ respectively. Notice that the 
abscissas correspond to the fraction of training patterns 
per input space dimension. For each value of $\ell$, 
averages were performed over a large enough number of 
different teachers and training sets to get error bars 
smaller than the symbols' size in the figures. Experiments 
were carried out for $n=50$ and $n=20$. The corresponding 
feature space dimensions $n(n+1)$ are $2550$ and $420$ 
respectively. As both show the same trends, only results for 
$n=50$ are presented on the figures. For the sake of comparison 
with the theoretical results, we characterize the actual 
SVM by $\Delta$, its (finite size) inflation factor 
(\ref{eq.delta}), and $\sigma^2$, the variance (\ref{eq.sigma2}) 
of the components in the $\sigma$-subspace. Since $n=50$, we 
have $\Delta=25.5$ and $\sigma^2 = 1.960784 a^2$,
that is $\sigma^2 = 1.960784$ for the non-normalized mapping 
and $\sigma^2 = 0.039216$, for the normalized one. 

The theoretical values of $Q$, the fraction of 
squared student weights in the 
$\sigma$-subspace, and the teacher-student overlaps $R^u$ and 
$R^\sigma$, corresponding to the same classes of 
teachers as the experimental results are represented as 
lines on the same figures. The excellent 
agreement with the experimental data is striking, 
and gives an indication that the high order correlations of the 
features, neglected in the model, 
are indeed negligible. Therefore, we have $R^l=R^u$, 
and $R^q = R^\sigma$, so that from now on we can drop 
the indices $l,q$ and keep $u,\sigma$ used in 
the theoretical approach.

\begin{figure}
\centerline{\psfig{figure=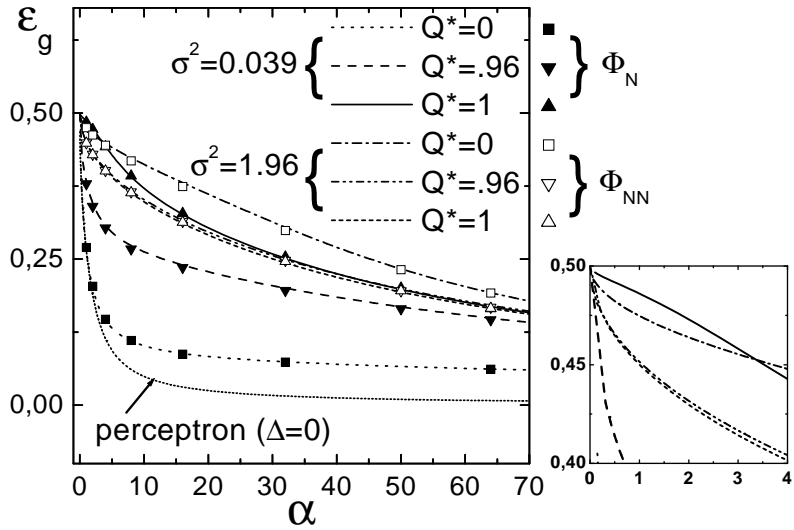,height= 8 cm}}
\caption{Learning curves of SVMs for different teacher 
rules $Q_*$. Definitions are the same as in figure 
{\protect {\ref{fig.Q*0}}}. The inset is an 
enlargement of the small $\alpha$ region.}
\label{fig.epsg}
\end{figure}

Fig. \ref{fig.Q*0} corresponds to a purely linear 
teacher ($Q_*=0$), i.e. to a quadratic SVM learning a 
rule linearly separable in input space. As in this case
$R^\sigma=0$ because ${\bf w}_*^\sigma=0$, only $R^u$ and $Q$ 
are represented. Conversely, in the case of a purely 
quadratic rule, $Q_*=1$, represented on fig. \ref{fig.Q*1}, 
$R^u=0$. Notice that $Q \rightarrow Q_*$ for 
$\alpha \rightarrow \infty$ in both cases, meaning that the 
student learns in which subspace lies the teacher. 
Correspondingly, the normalized teacher-student overlap 
increases smoothly to $1$. However, the pace at which 
these quantities reach their asymptotic limit depends 
on the features normalization factor, and it is their 
combination given by eq. (\ref{eq.R}) that determines 
the generalization error $\epsilon_g$, represented on 
fig. \ref{fig.epsg}. Clearly, if $Q_*=0$, i.e. if the 
problem is linearly separable in input space, the normalized 
mapping has lower $\epsilon_g$ at finite $\alpha$. 
Conversely, if the discriminating surface is purely 
quadratic in input space, the non-normalized mapping 
gives better results. In the asymptotic limit 
$\alpha \rightarrow \infty$ the mapping's normalization 
becomes irrelevant, as in both cases the generalization 
error vanishes.

Fig. \ref{fig.Q*51/53} shows the results corresponding 
to the isotropic teacher, 
$Q_*=Q_{*;\,iso} \equiv \Delta/(1+\Delta)$. For 
$\Delta=25.5$ we have $Q_{*;\,iso}=0.962$. 
A particular case of such a teacher, considered 
in \cite{DOS,YO}, has all its weight components 
of equal absolute value $|w_{*,\nu}|=|w_{*, \nu \mu}|=1$. 
Finally, the results corresponding to a general 
rule, with $Q_*=0.5$, are shown in fig. \ref{fig.Q*05}. 

Notice that, irrespective of the mapping, the overlaps 
$R^u$ and $R^\sigma$ present different behaviours, as 
the latter increases much slower than the former. 
This reflects the fact that, as the number of quadratic 
components scales like $n \Delta$, a number of examples 
of the order of $n \Delta$ are needed to learn them. Thus, 
as a function of $\alpha$, the linear components are 
learned first. Indeed, $R^u$ reaches a value close to $1$ 
with $\alpha \sim O(1)$ while $R^\sigma$ 
needs $\alpha \sim O(\Delta)$ to reach similar values. We 
call this general trend {\it hierarchical learning}. 

As the generalization error depends on $R^u$ 
and $R^\sigma$ through the combination (\ref{eq.R}), the 
signature of hierarchical learning is present on the learning 
curves $\epsilon_g$ corresponding to the different rules, 
plotted against $\alpha$ on fig. \ref{fig.epsg}. 
The performance obtained 
with the normalized mapping is better the smaller the value 
of $Q_*$. The non-normalized mapping shows the opposite 
trend: its performance for a purely linear teacher ($Q_*=0$) is 
extremely bad, but it improves for increasing values of 
$Q_*$ and slightly overrides that of the normalized 
mapping in the case of a purely quadratic teacher.
 
These results reflect the competition on learning the 
anisotropically distributed features. The more the features 
are compressed, the more difficult the learning task. In the case of 
rules with $Q_* \ll 1$, the linear components carry 
the most significant information. In those cases, it 
is advantageous to use the normalized mapping, 
which has the $\sigma$-components compressed 
($\sigma^2=0.039$) with respect to the 
{\it u}-components, which have unit variance. 

The non-normalized mapping has $\sigma^2=1.96$, meaning 
that the compressed components are those of the 
{\it u}-subspace. This mapping is better whenever most 
of the information is contained in the $\sigma$-subspace, 
which is the case for teachers with large $Q_*$ and, in 
particular, with $Q_*=1$. In this last case, the 
linear components only introduce 
noise that hinders the learning process. As the 
number of linear components is much smaller than the 
number of quadratic ones, their pernicious effect is 
expected to be more conspicuous the smaller the value 
of $\Delta$. 

\begin{figure}
\centerline{\psfig{figure=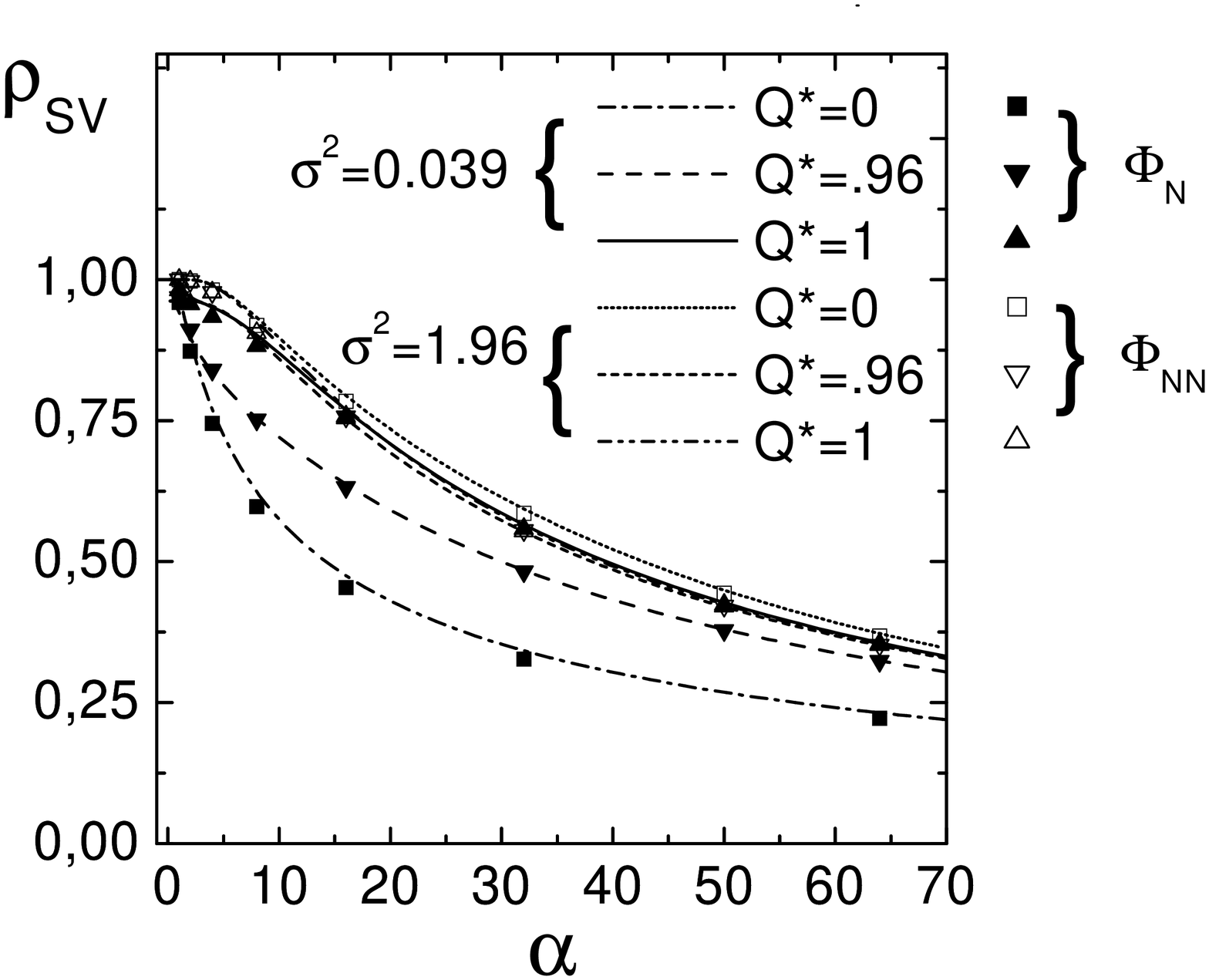,height= 8 cm}}
\caption{Fraction of learning patterns that belong to the subset of Support Vectors.}
\label{fig.rhoSV}
\end{figure} 

Finally, for the sake of completeness, the fraction of 
support vectors $\rho_{SV} \equiv \ell_{SV}/\ell$, where 
$\ell_{SV}$ is the number of training patterns with maximal 
stability, is represented on figure \ref{fig.rhoSV}. 
Notice that, although these curves present qualitatively 
the same trends as $\epsilon_g$, they constitute a very 
loose bound to the latter. Since the student's 
weights can be expressed as a linear combination of 
SVs \cite{Vapnik}, this result is of practical interest. 
It shows that increasing the number of training patterns 
does not increase the typical number of non-vanishing 
coefficients that have to be determined using the quadratic 
minimization program: although at low $\alpha$ most of 
the training patterns are expected to be support vectors, 
this fraction decreases smoothly with $\alpha$. Notice that 
$\alpha$ in our approach is the number of training patterns 
per {\it input} space dimension. Thus, even if $\rho_{SV}$ 
may seem large, this is not so in regard to the dimension 
of the feature space. In our simulations for example, 
$\alpha=50$ is of the order of the VC-dimension of the 
feature space. However, from fig. \ref{fig.rhoSV} we expect 
that less than half of the training patterns be SVs. 

\section{Discussion}
\label{sec.discussion}

In order to understand the results obtained in the 
previous section, we first consider the relative 
behaviour of $R^u$ and $R^\sigma$, deduced from 
the theoretical approach. 
If $\Delta_*^\sigma \ll \Delta$, which is the case 
for sufficiently small $Q_*$, we get that $R^\sigma \ll R^u$. 
This means that the quadratic components are more difficult 
to learn than the linear ones. On the other 
hand, if the teacher lies mainly in the quadratic subspace, 
$\Delta_*^\sigma \gg \Delta$ and $R^\sigma > R^u$. 
The crossover between these different behaviours occurs at 
$\Delta_*^\sigma = \Delta$, for which $R^\sigma = R^u$. 
For $n=50$, this 
arises for $Q_{*}=0.998$ for the normalized mapping and 
for $Q_{*}=0.929$ for the non-normalized one. 
In the particular case of the isotropic teacher and the 
non-normalized mapping, $Q_{*;\,iso} > 0.929$, so that 
$R^\sigma > R^u$, as shown on figure \ref{fig.Q*51/53}.

These considerations alone are not sufficient to understand 
the behaviour of the generalization error, which depends on the 
weighted sum of $R^\sigma$ and $R^u$ (see equation 
(\ref{eq.R})). The behaviour of $R^\sigma$ and $R^u$ at 
small $\alpha$ is useful to understand the 
onset of hierarchical learning. In the limit 
$\alpha \rightarrow 0$, we find that 
$Q \simeq \Delta \sigma^2 / (\Delta \sigma^2 +1)$ 
to leading order in $\alpha$. This result may be understood 
with the following simple argument: if 
there is only one training pattern, clearly it is a 
SV and the student's weight vector is proportional to 
it. As a typical example has $n$ components of 
unit length in the {\it u}-subspace and $n \Delta$ components 
of length $\sigma$ in the $\sigma$-subspace, we have 
$Q=n\Delta \sigma^2 / (n \Delta \sigma^2 +n)$. With the 
normalized mapping, $\lim_{\alpha \rightarrow 0} Q = 1/2$. 
In the case of the non normalized one $\lim_{\alpha \rightarrow 0} Q  
= (2 \Delta - 1)/2 \Delta$, which depends on the inflation factor 
of the SVM. In this limit, we obtain:

\begin{eqnarray}
\kappa &\simeq& 
\frac{1+\sigma^2 \Delta}{\sqrt{1+\sigma^4 \Delta}} \frac{1}{\sqrt{\alpha}}, \\
R^u &\simeq& \sqrt{\frac{2}{\pi}} \frac{1}{\sqrt{1+\Delta_*^\sigma}} 
\sqrt{\alpha}, \\
R^\sigma &\simeq& \sqrt{\frac{2}{\pi}} \, \sqrt{\frac{\Delta_*^\sigma}{1+\Delta_*^\sigma}} \, 
\sqrt{\frac{\alpha}{\Delta}}.
\end{eqnarray}

\noindent Therefore, $R \sim \sqrt{\alpha}$ for 
$\alpha \ll 1$, as for the simple 
perceptron MMH \cite{GG}, but with a prefactor that depends 
on the mapping and the teacher. 

In our model, we expect that hierarchical learning correspond to 
a fast increase of $R$ at small $\alpha$, mainly
dominated by the contribution of $R^u$. As in the limit 
$\alpha \rightarrow 0$,

\begin{equation}
R \simeq \frac{R^u + R^\sigma \sqrt{\sigma^4 \Delta \Delta_*^\sigma}}
{\sqrt{1+\sigma^4 \Delta} \sqrt{1+\Delta_*^\sigma}},
\end{equation}

\noindent we do not expect to have hierarchical learning if  
$\sigma^4 \Delta \gg 1$, as in that case mainly 
$R^\sigma$ contributes to $R$. This 
is what happens with the non normalized mapping. On the other hand, 
hierarchical learning takes place if 
$\sigma^4 \Delta \ll 1$ and $\Delta_*^\sigma \lesssim 1$. 
The first condition establishes a constraint on the mapping, 
which is only satisfied by the normalized one. The second 
condition, that ensures that $R^\sigma < R^u$ holds, gives 
the range of teachers for which this hierarchical 
generalization takes place. Under these conditions, 
$R$ grows fast and the contribution 
of $R^\sigma$ is negligible because it is weighted 
by $\sqrt{\sigma^4 \Delta \Delta_*^\sigma}$. The effect 
of hierarchical learning is more important the smaller 
$\Delta_*^\sigma$. The most dramatic effect arises for $Q_*=0$, 
i.e. for a quadratic SVM learning a linearly separable rule.  

Notice that if the normalized mapping is used, the condition 
$\Delta_*^\sigma \lesssim 1$ implies that $Q_* < Q_{*;\,iso} \equiv
\Delta/(1+\Delta)$, where $Q_{*;\,iso}$ corresponds to the isotropic 
teacher. A straightforward calculation shows that if the teachers 
are drawn at random on the surface of the hypersphere in feature 
space, the distribution of $Q_*$ is highly non symmetric, with a 
maximum at the value of $Q_{*;\,iso}$, that depends on $n$. The 
fraction of teachers with $Q_* < Q_{*;\,iso}$ is smaller than 
$1/2$. For example, only $47.5 \%$ of the teachers satisfy this 
constraint if $n=50$. When $n \rightarrow \infty$, the distribution 
becomes $\delta(Q_* - Q_{*;\,iso})$, and $Q_{*;\,iso}$ tends to the 
median, meaning that in this limit, only about $50 \%$ of the 
teachers give rise to hierarchical learning when using 
the normalized mapping.   

In the limit $\alpha \rightarrow \infty$, all the 
generalization error curves converge to the same asymptotic 
value as the simple perceptron MMH learning in the feature 
space, namely $\epsilon_g=0.500489 (1+\Delta)/ \alpha$, 
independently of $\sigma$ and $Q_*$. Thus, $\epsilon_g$ vanishes 
slower the larger the inflation factor $\Delta$. 

\begin{figure}
\centerline{\psfig{figure=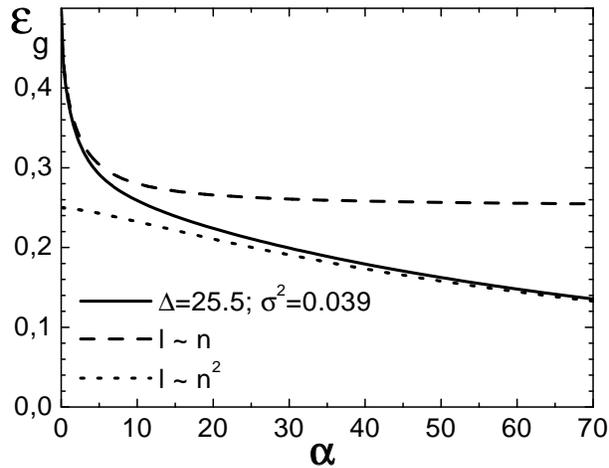,height= 7 cm}}
\caption{Generalization error of a SVM corresponding to 
different thermodynamic limits. See the text for the definition 
of $\alpha$ in each regime.}
\label{fig.thermolim}
\end{figure}

Since the inflation factor $\Delta$ of the SVM feature space 
in our approach is a free parameter, it does not diverge
in the thermodynamic limit $N \rightarrow \infty$ . As a 
consequence, the two scaling regimes for $\epsilon_g$ give rise 
to a simple crossover between a fast decrease 
at small $\alpha$ followed by a slow decrease at 
large $\alpha$. The results of Dietrich et al. \cite{DOS} 
for the {\it normalized} mapping, that corresponds to $\sigma^2 
\Delta=1$ in our model, can be deduced by taking appropriately 
the limits before solving our saddle point equations. 
The regime where the number of training patterns $\ell= \alpha n$ 
scales with $n$, is straightforward. It is obtained, 
within our approach, by taking the limit $\sigma \rightarrow 0$ 
and $\Delta \rightarrow \infty$ keeping 
$\sigma^2 \Delta=1$ and $\alpha$ finite. The 
regime where the number of training patterns 
$\ell= \alpha n$ scales with $n \Delta$, the number of quadratic 
features, is obtained by keeping 
$\tilde \alpha \equiv \alpha/(1+ \Delta)$ finite whilst taking, 
here again, the limit $\sigma \rightarrow 0$, 
$\Delta \rightarrow \infty$ with $\sigma^2 \Delta=1$. 
The corresponding curves are represented on figure 
\ref{fig.thermolim} for the case of an isotropic teacher. 
In order to compare with our results at 
finite $\Delta$, the regime where $\tilde \alpha$ is 
finite is represented as a function of $\alpha = 
(1+\Delta) \tilde \alpha$ using the value of $\Delta$ 
corresponding to $n=50$, namely, 
$\Delta = 25.5$. In the same figure we represented the 
generalization error $\epsilon_g$ 
obtained with our model using the parameter 
values $\sigma^2=0.039$ and $\Delta=25.5$.

These results, obtained for quadratic SVMs, are easily 
generalizable to higher order polynomial SVMs, as is the 
case with the approach of \cite{DOS}. 
We expect a cascade of hierarchical decreasings 
of the generalization error as a 
function of $\alpha$, in which successively more and 
more compressed features are 
learned. 

\section{Conclusion}
\label{sec.conclusion}

We introduced a model that clarifies some aspects of the 
generalization properties of polynomial Support Vector 
Machines (SVMs) in high dimensional feature spaces. To 
this end, we focused on quadratic SVMs. The quadratic 
features, which are the pairwise products of input 
components, may be scaled by a {\it normalizing factor}. 
Depending on its value, the generalization 
error presents very different behaviours in the 
thermodynamic limit \cite{DOS,BG}. 

We showed that any finite size quadratic SVM may 
be characterized by two 
parameters: $\Delta$ and $\sigma$. The 
{\it inflation factor} $\Delta$ is the 
number of quadratic features relative to 
the number of input  features, and is proportional 
to the input space dimension $n$. The {\it variance} 
$\sigma$ of the quadratic features is related to 
the normalizing factor. Usually, either
$\sigma \sim 1/\sqrt{n}$ (normalized mapping) or 
$\sigma \sim 1$ (non normalized mapping). 
In previous studies, not only the input space dimension 
diverges in the thermodynamic limit $n \rightarrow \infty$, 
but also $\Delta$ and $\sigma$ are correspondingly scaled.

In our model, the proportion of quadratic features 
$\Delta$ and their variance $\sigma$ are considered as 
parameters characterizing the (finite size) SVMs. Since 
we keep them constant when taking the thermodynamic limit, 
we can study the learning properties of actual SVMs with 
finite inflation ratios and normalizing factors, as a 
function of $\alpha \equiv \ell/n$, where $\ell$ is 
the number of training examples. 

Our theoretical results were obtained neglecting the 
correlations among the quadratic features. Indeed, 
their effect does not seem to be important, as the 
agreement between our computer experiments with actual 
SVMs and the theoretical predictions is excellent. This 
approximation was also shown to give good predictions in other 
similar problems \cite{DOS,YO}, but further investigations 
are needed to establish rigorously the conditions of 
its validity.

We find that the generalization error $\epsilon_g$ 
depends on the type of rule to be inferred through 
$Q_*$, the (normalized) teacher's squared weight 
components in the quadratic subspace. If the normalized 
mapping is used and $Q_*$ is small enough, the behaviour 
of $\epsilon_g$ at small $\alpha$ is dominated by 
the high rate learning of the linear components. On 
increasing $\alpha$, there is a crossover to a regime 
where the decrease of $\epsilon_g$ becomes much slower. 
This crossover becomes smoother for increasing values 
of $Q_*$, and this effect of 
hierarchical learning disappears for large enough $Q_*$. 
If the limits 
$\Delta \sim n \rightarrow \infty$ 
and $\sigma^2 \sim 1/n \rightarrow 0$
are taken together with the thermodynamic limit, the 
hierarchical learning effect gives rise to the different 
scaling regimes, corresponding to $\ell \sim n$ or 
$\ell \sim n^2$, described previously \cite{YO,DOS}.

On the other hand, if the features are not normalized, 
the contributions of both the linear and the quadratic 
components to $\epsilon_g$ are of the same order, and 
there is no hierarchical learning at all. 
For $Q_* > Q_{*;\,iso}$, which 
corresponds to the isotropic teacher, the non normalized 
mapping has a slightly smaller 
generalization error than the normalized one. 

It is worth remarking that if the rule to be learned 
allows for hierarchical learning, the generalization 
error of the normalized mapping is much smaller than 
that of the non normalized one. In fact, 
the teachers corresponding to such rules are those 
with $Q_* \lesssim Q_{*;iso}$, where $Q_{*;\,iso}$ 
corresponds to the isotropic teacher, the one 
having all its weights components equal. For the others, both 
the normalized mapping and the non normalized one 
present similar performances. If the weights of the 
teacher are selected at random on a hypersphere 
in feature space, the most probable 
teachers have precisely $Q_*=Q_{*;\,iso}$, and the fraction 
of teachers with $Q_* \leq Q_{*;\,iso}$ represent of the order of 
$50\%$ of the inferable rules. Thus, from a practical 
point of view, without having any prior knowledge 
about the rule underlying a set of examples, the 
normalized mapping should be preferred.
 
\section*{Acknowledgements}

It is a pleasure to thank Arnaud Buhot for a careful 
reading of the manuscript, and Alex Smola for 
providing us the Quadratic 
Optimizer for Pattern Recognition program \cite{AS}. 
The experimental results were obtained with the Cray-T3E 
computer of the CEA (project 532/1999). 

We thank the ZIF (Bielefeld), where the last version of 
the paper was revised, for its kind hospitality in the frame of the 
workshop ``The Sciences of Complexity''.

SR-G acknowledges economic support from the EU-research contract 
ARG/B7-3011/94/27. MBG is member of the CNRS.

\section*{Appendix}

The average (\ref{eq.lnZ}) within the replica approach can be 
expressed as a function of several order parameters, whose 
values have to minimize the free energy of the system. Among 
these, the  overlaps of the weights corresponding to 
different replicas, $a,b$, 

\begin{eqnarray}
q^u_{ab} &=& \sum_{\nu=1}^n \overline{\langle w_{a;\nu} w_{b;\nu}\rangle} 
/\tilde n,  \nonumber \\ 
q^\sigma_{ab} &=& \sum_{\nu=n+1}^{\tilde n}  
\overline{\langle w_{a;\nu} w_{b;\nu}\rangle} /\tilde n. 
\end{eqnarray}

As the tasks considered are learnable, the solution that 
minimizes the cost function (\ref{eq.cost}) with maximal 
$\kappa$ is unique. Thus, in the following we may assume that 
replica symmetry holds. Then, $q_u^{ab}=q_u$, 
$q_\sigma^{ab}=q_\sigma$ for all $a,b$, and the saddle 
point equations corresponding to the extremum of the 
free energy for the MMH are:

\begin{eqnarray}
\label{eq.sp1}
2 \frac{\alpha}{\Delta} \Delta^\sigma I_1 &=& (1-(R^\sigma)^2) \,
\frac{(x + \Delta^\sigma)^2}{1+\Delta^\sigma}, 
\\
\label{eq.sp2}
2 \frac{\alpha}{\Delta}  I_2 
&=& \sqrt{\frac{1+\Delta_*^\sigma}{\Delta_*^\sigma}} \, R^\sigma \,
\frac{x + \Delta^\sigma}{\sqrt{\Delta^\sigma (1+\Delta^\sigma)}}, 
\\
\label{eq.sp3}
2 \frac{\alpha}{\Delta} Q (1-\sigma^2) I_3 &=& 
\left(1- x \frac{1-(R^\sigma)^2}{1-(R^u)^2}\right) \, 
\frac{x + \Delta^\sigma}{1+\Delta^\sigma},
\\
\label{eq.sp4}  
\frac{(R^\sigma)^2}{1-(R^\sigma)^2}  &=& \frac{\Delta_*^\sigma}{\Delta} \frac{(R^u)^2}{1-(R^u)^2},
\\
\label{eq.sp5}
\frac{R^u}{R^\sigma} &=& x \frac{\Delta}{\sqrt{\Delta^\sigma \Delta_*^\sigma}}.
\end{eqnarray}

\noindent where $\Delta^\sigma$ and $\Delta_*^\sigma$ are 
defined by eq. (\ref{eq.deltas}), $Q_*$, $Q$, $R^u$ and $R^\sigma$ 
by equations (\ref{eq.Q*}) to (\ref{eq.Rsigma}), and $x \equiv lim_{\kappa 
\rightarrow \kappa_{max}} (1-q_u/(1-Q))/(1-q_\sigma/Q)$. The 
integrals in the left hand side of equations 
(\ref{eq.sp1}-\ref{eq.sp3}) are

\begin{eqnarray}
I_1 &=& \int_{-\tilde \kappa}^\infty Dt \, (t+{\tilde \kappa})^2 \, 
H\left(\frac{t R}{\sqrt{1-R^2}}\right),
\\
I_2 &=& \frac{1}{\sqrt{2 \pi}} \left[\frac{\sqrt{1-R^2} 
\exp(-{\tilde \kappa^2}/(2 (1-R^2)))} {\sqrt{2 \pi}} + 
{\tilde \kappa} H\left(\frac{-\tilde \kappa}{\sqrt{1-R^2}}\right)\right],
\\
I_3 &=& \int_{-\tilde \kappa}^\infty Dt \, \tilde \kappa \, (t+{\tilde \kappa}) \, H\left(\frac{t R}{\sqrt{1-R^2}}\right),
\end{eqnarray}

\noindent with $Dt \equiv dt \, \exp{(-t^2/2)}/\sqrt{2 \pi}$, 
$H(x)=\int_x^\infty Dt$, $R$ is given in eq. (\ref{eq.R}), and 

\begin{equation}
\tilde \kappa = \frac{\kappa_{max}}{\sqrt{(1-Q) (1+\Delta^\sigma)}}.
\end{equation}

After solving the above equations for $Q$, $R^u$, $R^\sigma$, 
$x$ and $\tilde \kappa$, they allow to determine $\rho_{SV}$ and 
$\epsilon_g$ through (\ref{eq.rho}) and (\ref{eq.eg}).

\end{article}

\end{document}